\documentclass{article}

\usepackage{arxiv}

\usepackage[utf8]{inputenc} % allow utf-8 input
\usepackage[T1]{fontenc}    % use 8-bit T1 fonts
\usepackage{hyperref}       % hyperlinks
\usepackage{url}            % simple URL typesetting
\usepackage{booktabs}       % professional-quality tables
\usepackage{amsfonts}       % blackboard math symbols
\usepackage{nicefrac}       % compact symbols for 1/2, etc.
\usepackage{microtype}      % microtypography
\usepackage{graphicx}
\usepackage{natbib}
\usepackage{doi}
\usepackage{amsmath}

\newcommand{\N}{\mathcal{N}}

\newcommand{\cL}{\mathcal{L}}

\newcommand{\x}{\mathbf{x}}

\newcommand{\y}{\mathbf{y}}

\newcommand{\W}{\mathbf{W}}

\newcommand{\p}{\mathbf{p}}

\newcommand{\Var}{\text{Var}}

\title{Towards Reducing Aleatoric Uncertainty for Medical Imaging Tasks}

\author{ 
	Abhishek Singh Sambyal\\
	Department of Computer Science and Engineering\\
	Indian Institute of Technology Ropar\\
	\And
	
	Narayanan C Krishnan\\
	Department of Computer Science and Engineering\\
	Indian Institute of Technology Ropar\\
	\And
	Deepti R Bathula\\
	Department of Computer Science and Engineering\\
	Indian Institute of Technology Ropar\\
	\date{}
}

\begin{document}
\maketitle

\begin{abstract}
	In safety-critical applications like medical diagnosis, certainty associated with a model’s prediction is just as important as its accuracy. Consequently, uncertainty estimation and reduction play a crucial role.  Uncertainty in predictions can be attributed to noise or randomness in data (aleatoric) and incorrect model inferences (epistemic). While model uncertainty can be reduced with more data or bigger models, aleatoric uncertainty is more intricate. This work proposes a novel approach that interprets data uncertainty estimated from a self-supervised task as noise inherent to the data and utilizes it to reduce aleatoric uncertainty in another task related to the same dataset via data augmentation. The proposed method was evaluated on a benchmark medical imaging dataset with image reconstruction as the self-supervised task and segmentation as the image analysis task. Our findings demonstrate the effectiveness of the proposed approach in significantly reducing the aleatoric uncertainty in the image segmentation task while achieving better or on-par performance compared to the standard augmentation techniques.
\end{abstract}

% keywords can be removed

\section{Introduction}
\label{sec:intro}

Deep neural networks have achieved state-of-the-art performance in a variety of machine learning tasks. While prediction accuracy is an important measure characterizing the model's goodness, another critical factor contributing to the model's trust in many safety critical applications including medical imaging, is the prediction's uncertainty \cite{edlawate} \cite{surveydataaugmentation} \cite{TANNO2021117366}. This has motivated the development of many methods to estimate the prediction uncertainty.

Broadly, the uncertainty in the model's output stems from two sources - the inherent limitation of the data such as the absence of rich features, presence of noise, termed as aleatoric uncertainty, and the limitations in the learned model referred to as epistemic uncertainty. Aleatoric uncertainty, also known as \textit{Data uncertainty} is the uncertainty caused due to errors in the measurement, i.e., uncertainty arising from the intrinsic variability in the data. It measures the variation in the output of a model due to changes in the input. The aleatoric uncertainty is supposed to be irreducible for a specific dataset; however, incorporating additional features or improving the quality of the existing features can assist in its reduction.  Epistemic uncertainty, also known as \textit{Model uncertainty} captures uncertainty in the model parameters. This uncertainty can be reduced by increasing the training data size or model capacity.

Recently, many efforts have been directed towards reducing model uncertainty using data augmentation, bayesian inference, and ensembling \cite{weightunc-blundell15} \cite{kendall01NIPS2017_2650d608} \cite{simplescalable} \cite{bayesiansegnet_bmvc}. However, aleatoric uncertainty has not received its due attention.  As it originates from the data generation process, it cannot be explained by acquiring more data. Several factors contribute to this randomness/noise in medical images, including patient movement, different scanners, and partial volume effect. Although challenging, it is crucial to address this uncertainty in risk-sensitive applications like medical imaging to improve the robustness of the prediction against data noise when making critical decisions.

We propose a novel approach to reducing aleatoric uncertainty in one task (like segmentation) by leveraging the uncertainty estimates from another task (like reconstruction) performed on the same dataset. As a self-supervised task, image reconstruction provides a unique opportunity to judiciously estimate aleatoric uncertainty and interpret it as noise associated with the data.

\section{Methodology}
\label{sec:prelims}
Given a dataset $\mathcal{D}$, of $N$ paired training examples, $\mathcal{D} = \{{\x }_i, \y_i\}_{i=1}^N$, the goal is to learn a function $g$ (a segmentation task) to predict $\y_i$ given $\x_i$ along with reducing the aleatoric uncertainty in the prediction. We achieve this by first defining an auxiliary self-supervision task $f$ on $\x_i$ to estimate the inherent noise/variations in $\x_i$. We consider the vanilla reconstruction task as the self-supervision task. The estimated noise is then integrated into the learning process for task $g$ to reduce the aleatoric uncertainty.

\subsection{Modelling Uncertainty}
\label{sec:modellinguncertainty} 
The reconstruction based self-supervision task involves predicting $\x_i$ given the input $\x_i$ through an autoencoder ($f$, parameterized by $\psi$). Note that in the following discussion, due to the nature of the reconstruction task,  $\y_i = \x_i$. We wish to capture both the aleatoric and epistemic uncertainty in the output of the reconstruction task. We employ the dropout variational inference \cite{dropoutvi} (also known as Monte Carlo (MC) dropout) as an approximation to the posterior over the Bayesian Neural Network (BNN) to estimate these uncertainties.
% We employ the dropout variational distribution formulation of \cite{dropoutvi}, to estimate these uncertainties. 
Drawing the model parameters $\hat{\psi} \sim q(\psi)$ from the approximate posterior we obtain the output consisting of both $\hat{y}_{ij}$ and the aleatoric uncertainty $\hat{\sigma}(x_{ij})$. Assuming a Gaussian likelihood to model the aleatoric uncertainty induces the following minimization objective:
\begin{equation*}
\cL_\text{BNN}(\psi) = \frac{1}{N_i} \sum_{j=1}^{N_i} \frac{1}{2\hat{\sigma}(x_{ij})^2} ||y_{ij} - \hat{y}_{ij}||^2 + \frac{1}{2} \log \hat{\sigma}(x_{ij})^2
\end{equation*}
where, $\hat{y}_{ij}$ is the regressed value of pixel $j$ of image $\x_i$, $\hat{\sigma}$ is the noise observation parameter dependent on $x_{ij}$ for $\hat{\psi}$ and $N_i$ is the number of pixels in the image $\x_i$.\\
% equation 7.
The loss has two components - the residual error obtained through a stochastic sample, and an uncertainty regularization term. The aleatoric uncertainty is learned implicitly from the loss function. 
As suggested in \cite{kendall01NIPS2017_2650d608} for achieving numerical stability, we train the network to predict the log variance resulting in the following minimization function, $s_i:=\log \hat{\sigma}(x_{ij})^2$:
\begin{equation*}
\cL_\text{BNN}(\psi) = \frac{1}{N_i} \sum_{j=1}^{N_i} \frac{1}{2}{\exp(-s_i)} ||y_{ij} - \hat{y}_{ij}||^2 + \frac{1}{2} s_i
\end{equation*}
Thus, the predictive uncertainty for pixel $x_{ij}$ can be approximated using:
\begin{equation*}
\Var(y_{ij}) \approx \left(\frac{1}{T} \sum_{t=1}^{T}(\hat{y}_{ij})_{t}^2 - \left( \frac{1}{T}\sum_{t=1}^{T}(\hat{y}_{ij})_{t}\right)^2 \right) + \frac{1}{T}\sum_{t=1}^{T}(\hat{\sigma}_{ij})_{t}^2
\end{equation*}
where, $(\hat{y}_{ij})_{t=1}^T$ and $(\hat{\sigma}_{ij})_{t=1}^T$ are the $T$ sampled outputs for randomly masked weights $\hat{\psi} \sim q(\psi)$. The first and the second terms of the summation correspond to the epistemic and aleatoric uncertainties respectively.

\subsubsection{Modelling Segmentation Task Uncertainty}
\label{sec:methodology}
Segmentation is a pixel-level classification task. For each pixel \textit{j}, we predict the class label and the uncertainty in the prediction. We estimate the \textit{aleatoric uncertainty} using heteroscedastic classification neural network (NN) \cite{kendall01NIPS2017_2650d608}. The heteroscedastic classification NN, $g$ (parametrized by $\W$) predicts $k$-dimensional logit and uncertainty vectors for a $k$-class segmentation task. Assuming a Gaussian distribution over the logits, a sample logit vector can be obtained as shown in Eq.\ref{eq:classificationlogits}. The sampled vector is squashed with the softmax function to obtain the classification probabilities.

\begin{equation}\label{eq:classificationlogits}
    \hat{\y}_{ij}|\W \sim \mathcal{N}(g^{\W}_{ij}, (\boldsymbol{\sigma}^{\W}_{ij})^2); \text{\hspace{8pt}}\p_{ij} = \text{Softmax}(\hat{\y}_{ij})
\end{equation}
where, $g^{\W}_{ij}$ and $(\boldsymbol{\sigma}^{\W}_{ij})^2$ are the model output and variance. \textit{Epistemic uncertainty} of the probability vector $\p_{ij}$ is summarized by measuring the entropy. $\p_{ij}$ is approximated using Monte Carlo integration, which averages the softmax predictions for a given input over $T$ sampled masked weights $\{\hat{\W} \sim q(\W)\}_{t=1}^T$ where, at any given step $\W:=\hat{\W}$.

\subsection{Interpreting Aleatoric Uncertainty from Reconstruction Task as a Noise Model}
\label{ssec:interpretingdatauncertainty}
By definition, aleatoric uncertainty captures variations in the output due to changes/noise in the input. The choice of the reconstruction as the self-supervision task gives a unique interpretation to the aleatoric uncertainty estimated at every pixel- models the noise at the pixel location.  The statistics of the pixel intensities at each location can also be used to model this noise. However, modeling dependencies between neighboring pixels becomes a challenge. On the other hand, the reconstruction task inherently models the interaction between adjacent locations and, therefore, can provide a richer model characterizing the uncertainty of the pixel intensities at each location. More concretely, the aleatoric uncertainty $\sigma(x_{ij})$ for the reconstruction task is interpreted as the variance in the pixel intensities at the $j^{th}$ location for the image $\x_i$.

\subsection{Using the Noise Model to Reduce Aleatoric Uncertainty in the Segmentation Task}
Our previous interpretation of the aleatoric uncertainty allows us to explicitly account for the data noise when learning a different task using the same dataset. The heteroscedastic aleatoric uncertainty quantification from the reconstruction task provides variance in the pixel intensities at every location. We propose to augment the training data for the segmentation task by sampling pixel intensities from the noise model. Specifically, for every image, and at every pixel location, we sample the intensities from the Gaussian distribution $\N(\hat{x}_{ij}, \hat{\sigma}(x_{ij}))$, where, $\hat{x}_{ij}, \hat{\sigma}(x_{ij})$ are the outputs of reconstruction model. The image created through this sampling process is associated with the original image's ground truth. The augmented dataset is used to train the segmentation model. We hypothesize that this process reduces the aleatoric uncertainty in the segmentation task.

\section{Results and Discussion}
\label{sec:print}
The proposed aleatoric uncertainty reduction method is modeled as a data augmentation technique. Therefore, we compare our method with other standard data augmentation techniques, pixel-level augmentation (adding Gaussian noise) and structure-level augmentation (Full Augmentation) \cite{surveydataaugmentation}. We use \cite{kendall01NIPS2017_2650d608} as the framework for uncertainty estimation. 

We evaluate our method, quantitatively and qualitatively, on brain 
tumor segmentation (BraTS 2018) with $k=2$ classes (background versus whole tumor). We partitioned the dataset into train ($60\%$), validation  ($20\%$) and test set ($20\%$). As a pre-processing step, we applied intensity normalization to each MRI slice from each patient independently by subtracting the mean and dividing by the standard deviation of the brain region computed at the patient level. We cropped the input image from $240 \times 240$ to $188 \times 188$, removing the background pixels as much as possible. We have used modified UNet architecture with a dropout probability of 0.5 applied throughout the network for all the experiments. The same architecture is used for both reconstruction and segmentation tasks.  We used AdamW optimizer with a learning rate of $10^{-3}$ and weight decay of $10^{-2}$. During training, the learning rate is reduced by the factor of $0.1$ with a patience of $20$. In practice, we used Laplacian prior, as opposed to the Gaussian prior. The resulting loss applies an $L1$ distance on the residuals. Based on our experimentation, we found this to perform better than $L2$ loss for the reconstruction task \cite{kendall01NIPS2017_2650d608}. We used dice loss for the segmentation task.

By definition, the \textit{uncertainties} are defined over the predictions, but instead of just evaluating them in the tumor region, we also compare how different augmentations perform in the non-tumor area of the brain.
To evaluate the \textit{quality of the segmentation}, we used six performance metrics: dice, precision, recall, F1, Jaccard index, and specificity, each applied to the brain region. To compare the \textit{calibration} of the model, we used two calibration metrics: expected calibration error (ECE) and Brier score (loss). The results are presented in Table \ref{tab:metrics}. We can see that our method shows significantly less aleatoric uncertainty than all other augmentations and the baseline, although the full augmentation constitutes both pixel-level and structure-level augmentation \cite{surveydataaugmentation}. Our method has also outperformed the baseline and Gaussian noise augmentation on almost all the performance and calibration metrics.

\begin{figure*}[!ht]
  \centering
%   \centerline{\includegraphics[width=\textwidth]{images/idea_01_image.pdf}}
  \centerline{\includegraphics[width=.9\textwidth]{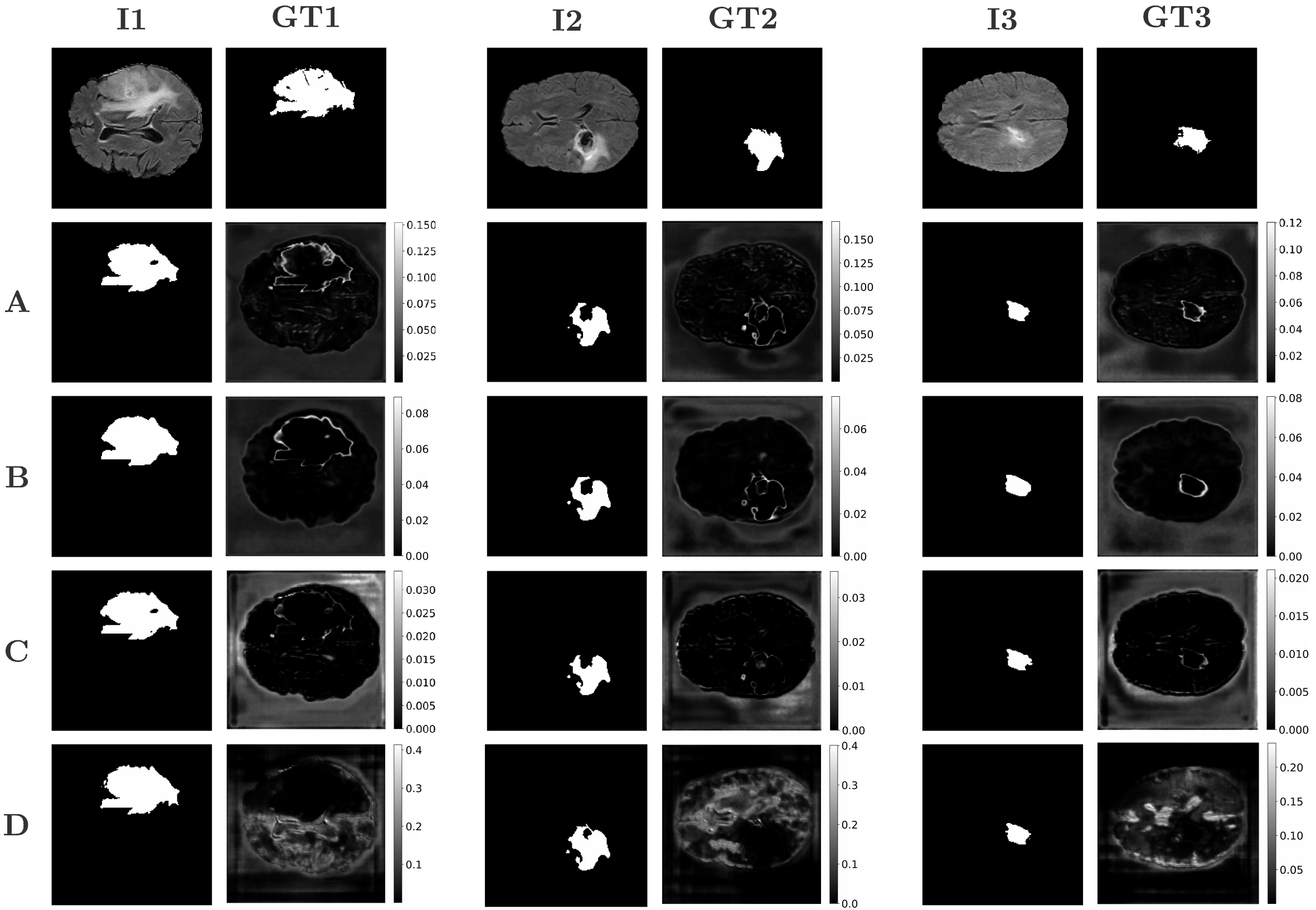}}
    % \centerline{\includegraphics[width=8.5cm]{example-image}}
  \caption{Comparison of prediction and aleatoric uncertainty maps. I1/I2/I3: Input Images, GT1/GT2/GT3: Ground Truth Images, A: Baseline, B: Gaussian, C: Ours, D: Full Augmentation. AI1 - DI1: Predicted Masks, AGT1 - DGT1: Aleatoric Uncertainty Maps. Similarly for I2/GT2 and I3/GT3.}
    \label{fig:uncertaintyimages}
\end{figure*}
In the aleatoric uncertainty maps, presented in Figure \ref{fig:uncertaintyimages}, we see higher uncertainty around the tumor boundary for the baseline and Gaussian noise-based augmentation. Our method shows minimum uncertainty compared to other augmentations (evident from the intensity scale). Although the full-augmentation uncertainty is also very less (vs. baseline/Gaussian), we see higher uncertainty in the region outside the tumor area.

We believe that the sampling process from the aleatoric uncertainty regions estimated by the reconstruction task helped to generate features, which assisted the segmentation model to reduce the aleatoric uncertainty significantly.

\begin{table*}[!ht]
\centering
\caption{Comparison between different implementations. i) Baseline: Segmentation model with an uncertainty estimation framework. (ii) Gaussian: Gaussian noise is added to the baseline model. (iii) Full Augmentation: Baseline model with four augmentations \cite{surveydataaugmentation} - (a) affine image transformations - scale, shear, rotate, vertical, horizontal flipping (b) elastic transformations. $\uparrow$: Higher is better, $\downarrow$: Lower is better; Best results shown in 
\textbf{bold}. Statistical difference between ours and best/2nd best: \textbf{**} $p < 0.001$ (highly statistically significant) \& $p > 0.05$ (statistically non-significant) shown in \textit{Italics}.}

\label{tab:metrics}
\resizebox{\textwidth}{!}{%
\begin{tabular}{|l | c  c  c | c |}
% \toprule
\hline
{} & Baseline &  Gaussian &      Ours &  Full Augmentation \\\hline
% \midrule
\hline
Aleatoric (Tumor) $\downarrow$ &  $0.01760 \pm 5.4e^{-5}$ &  $0.00773 \pm 1.7e^{-5}$ & **{\boldmath$0.00154 \pm 1.5e^{-6}$} &  $0.01361 \pm 3.6e^{-4}$\\\hline
% var\_dilated\ &  0.000054 &  0.000017 &  0.000002 &  0.000366 \\
Epistemic (Tumor) $\downarrow$ &  $0.08636 \pm 4.2e^{-3}$ &  $0.09399 \pm 5.9e^{-3}$ &  $0.08607 \pm 4.2e^{-3}$ &  **{\boldmath$0.07601 \pm 4.6e^{-3}$}\\\hline
% var\_dilated\ &  0.004241 &  0.005916 &  0.004234 &  0.004616 \\
Predictive (Tumor) $\downarrow$ &  $0.10397 \pm 5.0e^{-3}$ &  $0.10173 \pm 6.3e^{-3}$ & **{\boldmath$0.08761 \pm 4.3e^{-3}$} &  $0.08962 \pm 7.0e^{-3}$\\\hline
% var\_dilated\ &  0.005091 &  0.006374 &  0.004374 &  0.007089 \\
Aleatoric (Non-Tumor) $\downarrow$ &  $0.00369 \pm 1.4e^{-6}$ &  $0.00167 \pm 6.6e^{-7}$ & **{\boldmath$0.00050 \pm 1.6e^{-7}$} & $0.07141 \pm 7.3e^{-4}$\\\hline
% var\_brain\ &  0.000001 &  0.000001 &  0.000000 &  0.000733 \\
Epistemic (Non-Tumor) $\downarrow$ &  $0.00051 \pm 1.4e^{-6}$ &  $0.00066 \pm 1.4e^{-6}$ &  $0.00043 \pm 7.3e^{-7}$ &  **{\boldmath$0.00039 \pm 8.5e^{-7}$} \\\hline
% var\_brain\ &  0.000001 &  0.000001 &  0.000001 &  0.000001 \\
Predictive (Non-Tumor) $\downarrow$ &  $0.00421 \pm 2.7e^{-6}$ &  $0.00233 \pm 2.1e^{-6}$ &  **{\boldmath$0.00093 \pm 9.3e^{-7}$} &  $0.07181 \pm 7.3e^{-4}$ \\\hline\hline
% var\_brain\ &  0.000003 &  0.000002 &  0.000001 &  0.000735 \\
Dice $\uparrow$  &  $0.784 \pm 0.083$ &  $0.784 \pm 0.081$ &  $\mathit{0.790 \pm 0.081}$ &  {\boldmath$0.801 \pm 0.079$}  \\\hline
% var\_a\_scores\_Dice                        &  0.083593 &  0.081093 &  0.081504 &  0.079492 \\
Precision $\uparrow$ &  $0.603 \pm 0.188$ &  $0.573 \pm 0.172$ &  {\boldmath$\mathit{0.618 \pm 0.186}$} &  $0.606 \pm 0.184$ \\\hline
% var\_a\_scores\_Precision                             &  0.188724 &  0.172369 &  0.186343 &  0.184061 \\
Recall(Sensitivity/TPR) $\uparrow$    &  $0.531 \pm 0.168$ &  {\boldmath$0.559 \pm 0.175$} &  $\mathit{0.535 \pm 0.168}$ &  $0.552 \pm 0.177$ \\\hline
% varRecall(Sensitivity/TPR)                &  0.168648 &  0.175929 &  0.168843 &  0.177025 \\
F1 $\uparrow$ &  $0.543 \pm 0.165$ &  $0.544 \pm 0.161$ &  $\mathit{0.548 \pm 0.165}$ &  {\boldmath$0.558 \pm 0.169$} \\\hline
% var\_a\_scores\_F1                        &  0.165022 &  0.161812 &  0.165692 &  0.169203 \\
Jaccard $\uparrow$  &  $0.716 \pm 0.091$ &  $0.714 \pm 0.088$ &  $\mathit{0.723 \pm 0.089}$ &  {\boldmath$0.736 \pm 0.086$} \\\hline
% var\_a\_scores\_Jaccard                   &  0.091274 &  0.088817 &  0.089847 &  0.086949 \\
Specificity(TNR) $\uparrow$ &  $0.9940 \pm 1.3e^{-4}$ &  $0.990 \pm 2.4e^{-4}$ &  **{\boldmath$0.9943 \pm 1.1e^{-4}$} &  $0.992 \pm 1.8e^{-4}$ \\\hline\hline
% var\_pecificity(TNR)                      &  0.000130 &  0.000245 &  0.000114 &  0.000185 \\
ECE $\downarrow$ &  $0.0161$ &  $0.0168$ &  $0.0150$ &  {\boldmath$0.0146$} \\\hline
% MCE    &       $\downarrow$                                      &  0.3692 &  0.3144 &  0.3724 &  0.2951 \\\hline
Brier Score $\downarrow$ &  $0.0174$ &  $0.0183$ &  $0.0162$ &  {\boldmath$0.0157$} \\\hline
\bottomrule
\end{tabular}
}

\end{table*}

\section{Conclusion}

We propose a novel interpretation of aleatoric uncertainty estimated from an auxiliary self-supervised task as the noise or randomness inherent to the data and utilize it to reduce aleatoric uncertainty in other tasks performed on the same dataset. Experiments were performed on the benchmark BraTS dataset with image reconstruction as the self-supervised task and segmentation as the image analysis task. Data uncertainty estimated from reconstruction was used for data augmentation in segmentation by sampling images from the pixel predictive distribution. Our results show that the proposed approach significantly reduces aleatoric uncertainty in tumor segmentation compared to other standard augmentation methods. We further observe that the model's performance across many quantitative metrics is either better or on-par with other techniques, establishing it as a potentially reliable mechanism for addressing aleatoric uncertainty.

\section{Acknowledgements}
The support and the resources provided by `PARAM Sanganak Facility' under the National Supercomputing Mission, Government of India at the Indian Institute of Technology Kanpur is gratefully acknowledged.

\bibliographystyle{unsrtnat}
% \bibliography{references}  %%% Uncomment this line and comment out the ``thebibliography'' section below to use the external .bib file (using bibtex) 
\bibliography{my_ref}





% \end{thebibliography}

\end{document}